\documentclass[useAMS,usegraphicx,usenatbib,galley]{mn2e}
\usepackage{graphicx}

\def\msun{\ifmmode {\rm M_{\odot}}\else $\rm M_{\odot}$\fi}
\def\Msun{\ifmmode {\rm M_{\odot}}\else $\rm M_{\odot}$\fi}
\def\mearth{\ifmmode {\rm M_{\oplus}}\else $\rm M_{\oplus}$\fi}
\def\Mearth{\ifmmode {\rm M_{\oplus}}\else $\rm M_{\oplus}$\fi}
\def\Rearth{\ifmmode {\rm R_{\oplus}}\else $\rm R_{\oplus}$\fi}
\def\Ms{\ifmmode {M_s}\else $M_s$\fi}
\def\Mp{\ifmmode {M_p}\else $M_p$\fi}
\def\Rp{\ifmmode {R_p}\else $R_p$\fi}
\def\rearth{\ifmmode {\rm R_{\oplus}}\else $\rm R_{\oplus}$\fi}
\def\kep{{\it Kepler}}

\title[Mass Budget of Planet Forming Discs]
{The Mass Budget of Planet Forming Discs: Isolating the Epoch of Planetesimal Formation}
\author[Joan R. Najita and Scott J. Kenyon]{J. R. Najita$^{1,2}$\thanks{E-mail:
najita@noao.edu (JRN)} and S. J.  Kenyon$^{2}$ \\
$^{1}$National Optical Astronomy Observatory, 950 Cherry Avenue, Tucson, AZ. 85719, USA\\
$^{2}$Harvard-Smithsonian Center for Astrophysics, 60 Garden Street, Cambridge, MA 02138, USA}
\begin{document}

\date{}

\pagerange{\pageref{firstpage}--\pageref{lastpage}} \pubyear{2014}

\maketitle

\label{firstpage}

\begin{abstract}
The high rate of planet detection among solar-type stars argues
that planet formation is common. It is also generally 
assumed that planets form in protoplanetary discs like those 
observed in nearby star forming regions. On
what timescale does the transformation from discs to planets occur?
Here we show that current inventories of planets and protoplanetary discs 
are sensitive enough to place basic constraints on the timescale 
and efficiency of the planet formation process. A comparison of 
planet detection statistics and the measured solid reservoirs in 
T Tauri discs suggests that 
planet formation is likely already underway at the few Myr age of 
the discs in Taurus-Auriga, with a large fraction of solids having been converted 
into large objects with low millimeter opacity and/or sequestered 
at small disc radii where they are difficult to detect at 
millimeter wavelengths.

\end{abstract}


\section{Introduction}
\label{sec: intro}

Throughout the galaxy, nearly every young star is born with an opaque 
circumstellar disc of gas and dust \citep{haisch2001,mama2009,will2011}. 
Among T Tauri stars in the Taurus-Auriga dark cloud with ages of a few 
Myr \citep{luhman2010}, the discs have outer radii of 30--200~AU and 
masses of roughly $3 \times 10^{-4}$ to 0.25 times the mass of the 
central star \citep{and2013}.  By the time stars reach ages of 10--15 Myr, 
only a few show some evidence for large amounts of gas or dust 
\citep[see also][]{hernan2007,currie2009,kk2009,esp2014}.  
In the rest, the gas and dust have disappeared.

In current theories of star and planet formation, various physical processes
convert the gas and dust surrounding young stars into planetary systems
\citep[e.g.,][]{youdin2013}. 
In the popular core accretion picture, the dust grains in discs are first 
transformed into planetesimals, the building blocks of planets
\citep[e.g.,][]{gold1973,weiden1980,youdin2004a,rice2004,rice2006,
johansen2007,birn2010,youdin2011a,windmark2012,garaud2013}.
Planetesimals then grow collisionally to produce terrestrial planets and 
the solid cores of giant planets \citep[e.g.,][]{weth1993,pollack1996}. 
Throughout this agglomeration process, several mechanisms -- gas drag, 
type I migration, and scattering -- can cause large grains, planetesimals, 
or planets to migrate through the disc \citep{ada76,weiden1977a,malhotra1995,
rasio1996, ward1997,juric2008}. 
Currently, there is little agreement on the relative timing of growth 
and migration for known exoplanets observed close to their host stars.  
In one scenario, grains and planetesimals migrate inward and then grow 
into massive planets more or less {\it in situ} 
\citep[e.g.,][]{hansen2012,chiang2013,hansen2013}. Alternatively, 
massive planets might first grow farther away from the host star 
and then migrate or scatter inward 
\citep[e.g.,][]{ida2005,ida2008a,kk2008b,mann2010,raymond2014}.

Apart from the theoretical uncertainties, several observations support the 
notion that planets form in discs surrounding T Tauri stars.  The mass of 
the `minimum mass solar nebula' (MMSN) -- the minimum amount of material 
required to produce the planets in the solar system -- lies within the 
observed mass range of T Tauri discs with ages of a few Myr 
\citep{weiden1977b,hayashi1981,chiang2010,and2013}.
Many stars with ages of 10~Myr and older are surrounded by optically thin 
rings or discs of dusty debris \citep{wyatt2008,matthews2014}.  
For ensembles of stars 
with a broad range of ages, the properties (mass, temperature, and frequency) 
of dusty debris discs are consistent with the expected evolution of an 
ensemble of invisible 10--100~km objects left over from the formation 
of much larger planets \citep{kb2008,kb2010,kw2010}. 
Sensitive direct images of several of these systems reveal Jupiter mass 
planets \citep{marois2008,lagrange2010} which may sculpt the disc 
\citep[e.g.,][]{wyatt1999,wilner2002,stark2009,theb2012}.
Searches for planets using microlensing, radial velocity, and transit data
also detect an amazing diversity of planets orbiting older stars
\citep[e.g.,][]{how2010,sumi2011,how2012}.
The masses of these planetary systems lie within the known mass range of 
T Tauri discs. 
Thus, it seems obvious that exoplanets form out of T Tauri discs.

Despite this clear picture, the high incidence rates of planetary systems 
\citep{cum2008,gould2010,mayor2011,cassan2012,batalha2013,fressin2013,morton2013}   
now appear to challenge the ability of 
the reservoirs of solids in T Tauri discs 
to account for the solids bound 
up in planets and in the parent bodies of debris discs. 
When T Tauri disc 
masses were first measured in the 1990s, their masses were safely above 
the MMSN mass \citep{beck1993,oster1995}.
Debris discs and massive planetary systems were relatively rare.
Today, improved technology enables a more complete census of 
protoplanetary disc masses, 
with clear detections at 
very low solid masses, $\le$ 1--3~\mearth\ \citep[e.g.,][]{and2013}. 
We also now know that nearly all stars have one -- possibly more -- 
Earth-mass to Jupiter-mass planet 
\citep{gould2006,mayor2011,youdin2011b,cassan2012,petig2013,morton2013}. 
Moreover, in the three decades since the first discoveries of discs of small 
dust grains surrounding Vega and $\beta$ Pic \citep{aum1984,smith1984},
observations have revealed substantial infrared (IR) excesses associated 
with hundreds of normal main sequence stars and subgiants with ages ranging 
from a few Myr to 10~Gyr 
\citep[e.g.,][]{bac1993,hab2001,rieke2005,rhee2007,greaves2010a,eiroa2013,
bonsor2014,chen2014}.  

Taken together, the plethora of Earth-mass planets and very low mass 
T Tauri discs suggest a clear need to examine whether the solids available 
in known ensembles of T Tauri discs are adequate to explain the solids 
observed in planetary systems surrounding 10~Myr to 10~Gyr old stars.
Previous discussions of this issue focus on the ability of massive
T Tauri discs to explain the frequency of gas giant planets 
\citep[e.g.,][]{greaves2010b,vorobyov2011,will2012}. 
Given the theoretical inefficiency of planet formation, gas giants 
are expected to form in discs 
more than 2--3 
times the MMSN mass 
\citep[e.g.,][]{dod-rob2009}.
Because T Tauri discs of this mass are rarer 
than giant planets \citep{and2005, cum2008}, 
the assumed inefficiency of planet formation implies that 
planet formation begins before the T Tauri phase.

Recent advances in inventories of discs and exoplanets now allow 
us to infer the epoch of planet formation more directly. 
As we describe below, a comparison of the distributions of disc and 
exoplanet masses from $\sim$ 1~\Mearth\ to several $M_J$ leads to 
the stronger implication that planet formation is likely underway in 
class II discs {\it independent of assumptions about the efficiency 
of planet formation}. 

After reviewing current knowledge of the frequency and masses of planets 
and debris discs around nearby stars (\S2), we examine the most sensitive 
measurements of the mass available in discs surrounding young stars (\S3).  
Comparing the required mass with the available disc mass (\S4) demonstrates 
that the mass reservoirs in T Tauri discs match the frequency and masses of 
(i) planets within 1~AU of their host star,
(ii) gas giants at 1--10~AU, and
(iii) debris discs at $\ge$ 20~AU from the host star. 
However, the available mass in T Tauri discs is insufficient to explain 
the frequency of 
5--30~\mearth\ microlensing planets at 1--10~AU from the host star. 
While some factors -- measurement errors, planets with negligible 
core masses, and uncertainties in disc mass -- complicate our 
analysis (\S5.1), our approach is conservative (\S 5.2) 
and these issues are unlikely to change our conclusions (\S 6).

\section{The incidence rates of planets and debris discs}
\label{sec: planets}

To inventory the solid masses associated with mature 
planetary systems, we consider the frequency of 
planets and debris discs around old stars (age $\gg$ few Myr). 
Robust ground-based and space-based 
surveys paint a rich picture of planets close to 
(radial velocity and transit data) and far from 
(microlensing) the host star. Surveys with {\it IRAS}, 
{\it Spitzer}, and {\it Herschel} identify
debris discs surrounding stars over a range of ages. 
In this discussion,
we focus on results for FGK main sequence stars 
where the available methods yield fairly 
large ensembles of debris discs and planets.

\subsection{\bf Super-Earths and Neptunes:}

Planets in the 1--30~$\Mearth$ mass range are very common. 
In radial velocity studies with the HARPS spectrograph,
$\sim 54$\% of stars with FGK spectral types have a 
1--30~\mearth\ planet with an orbital period $P\le 100$\,d 
\citep{mayor2011}.  Roughly 70\% of these planets have 
masses of 1--10~\mearth.  Only a few percent have a giant 
planet more massive than $30~\mearth$.

Analyses of \kep\ data suggest similar results 
\citep{dong2013,petig2013,foreman2014}. For planets with
$P \le$ 100~d, the incidence rates are 20--30\% for 
1--2~\rearth\ planets and 20--40\% for 2--4~\rearth\ planets.
Using the planet mass-radius scaling 
for
planets in 
our Solar System ($\Mp/\Mearth = (\Rp/\Rearth)^{2.06}$,
Chiang \& Laughlin 2013), these two radius ranges correspond 
to the mass ranges 1--4\,$\Mearth$ and 4--17\,$\Mearth$.

These high incidence rates appear to continue to longer periods. 
Several analyses of low mass \kep\ planet candidates demonstrate 
that the incidence rate of planets is approximately constant with 
$\log P$ \citep{dong2013,petig2013,foreman2014}.
Among low mass planets with radii $\Rp \approx$ 1--4~\rearth, 
there are roughly half as many planets detected in the period
interval $P$ = 100--400~d as there are with $P \le$ 100~d. 

\begin{figure}
\includegraphics[width=17.5cm]{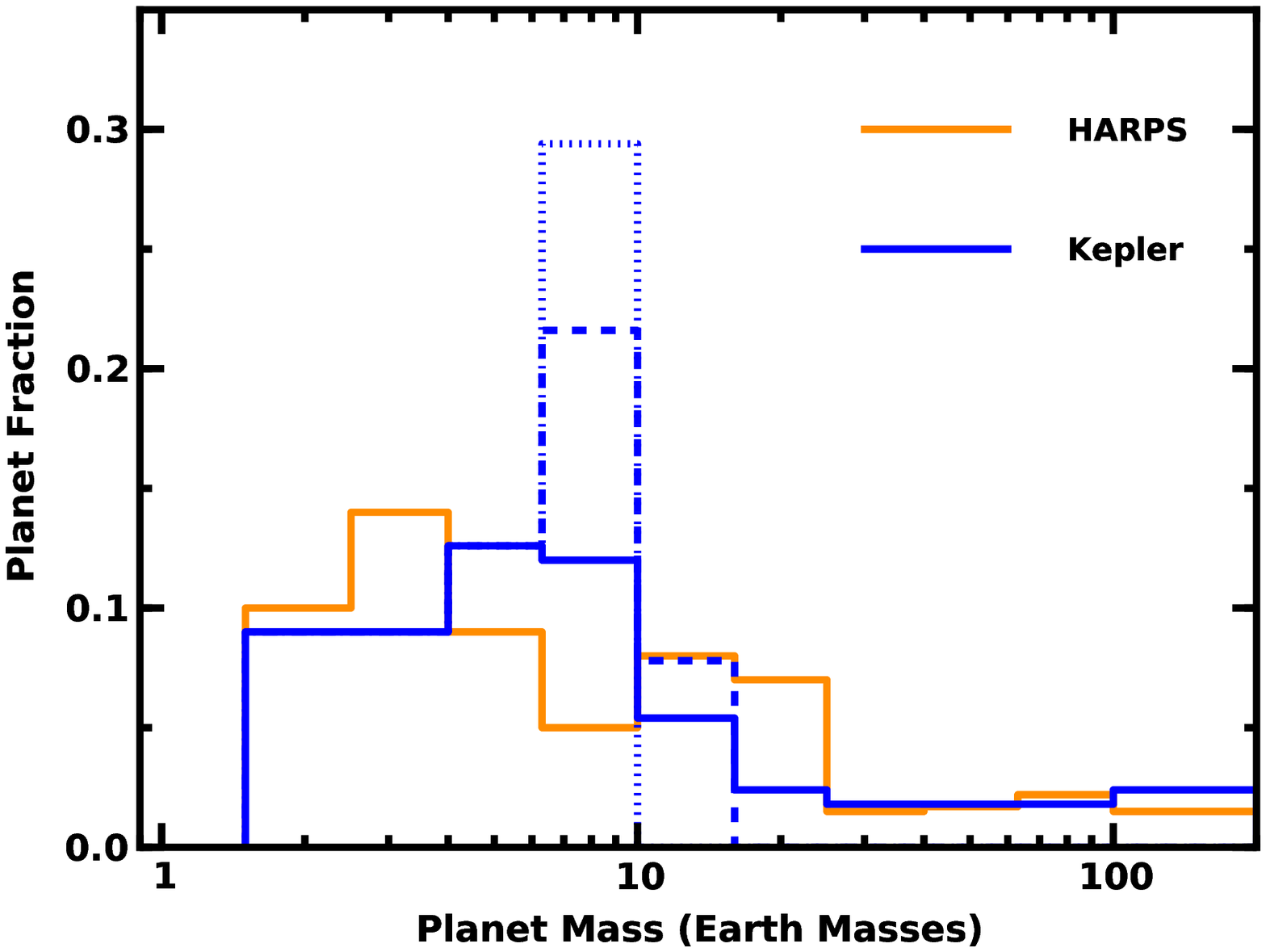}
\caption{
Differential incidence rates for exoplanets close to their host stars.
As summarized in the legend, solid lines show rates for 
(i) \kep\ planets with $P \le$ 125~d from this paper 
\citep[blue line; see also][]{dong2013}; and
(ii) HARPS planets with $P \le$ 100~d 
\citep[orange line;][]{mayor2011}.
Dotted (dashed) lines plot incidence rates for the 
solid mass in \kep\ planets assuming core masses of 
10~\mearth\ (7.5--12.5~\mearth).
\label{fig: exo-mass1}
}
\end{figure}

If we extrapolate the HARPS results with orbital period in the same way,  
the planet incidence rates for P $\le$ 400\,d from the HARPS 
survey are $\sim 27$\% (1.5--3~\mearth), $\sim 32$\% (3--10~\mearth), 
and $\sim 22$\% (10--30\,$\Mearth$).
The resulting total incidence rate of 81\% in the 
mass range 1.5--30\,$\Mearth$ is
similar to the \kep\ incidence rate of $\sim$ 75\% 
for the mass range 1--17~\mearth\ \citep{dong2013,foreman2014}.
Figure \ref{fig: exo-mass1} shows the differential incidence rate of 
planets as a function of mass from Mayor et al.\ (2011) after their 
correction for detection bias and assuming a cumulative incidence 
rate of 54\% for planets with $\Mp$ = 1.5--30~\mearth\ and $P \le $ 
100~d (solid orange curve).
The differential incidence rate generally increases with decreasing 
log $M$ from 30~\mearth\ to 3~\mearth.

To supplement these results, we consider several samples of transiting 
planets from the \kep\ {\it Space Telescope}. 
After downloading the \kep\ planet candidates from 
the first six \citep[Q1--Q6;][]{batalha2013} and first eight 
\citep[Q1--Q8;][]{burke2014} quarters of operation, we extracted candidates 
with signal-to-noise ratios of 8, orbital periods $P \le 125$~d, and 
host star effective temperatures $T_{\rm eff}$ = 5000--6500  
\citep[spectral types of F5--K2 in][see also Dong \& Zhu 2013]{kh1995}.  
From original samples of 2338 (3864) planets in the Q1--Q6 (Q1--Q8) data, 
this cut reduced the sample to 1582 (2495) planets orbiting 1251 (2067)
host stars.

To derive the incidence rates for \kep\ planets, we follow several simple 
steps.  For each planet, the mass $\Mp$ depends on the radius $\Rp$ 
\citep{liss2011,chiang2013}:
\begin{equation}
\Mp \approx \left ( { \Rp \over \rearth } \right)^{2.06} ~ \mearth.
\label{eq: m-planet}
\end{equation}
Although more detailed mass-radius relations are available
\citep[e.g.,][]{seager2007,enoch2012,lopez2013,weiss2013,weiss2014},
these often cover a small range in $\Rp$ or require knowledge of the 
host star metallicity.  Equation~(\ref{eq: m-planet}) generally 
provides a reasonable representation of these more detailed results
without introducing additional parameters.

Estimating the mass in solids for these planets requires an algorithm to
account for the mass in gas. In the core accretion theory for ice and gas
giants, small solids agglomerate into a protoplanet which accretes gas 
after reaching a mass $M_0$ \citep[e.g.,][]{pollack1996,rogers2011}.  
A simple estimate for $\Ms$, the mass in solids in every planet, is then
\begin{equation}
\Ms = \left\{
\begin{array}{lll}
M_p & \hspace{10mm} & M_p \le M_0 \\
M_0 & & M_p \ge M_0 \\
\end{array}
\right.
\label{eq: m-solid1}
\end{equation}
where $M_0$ is the typical mass of the solid core in a gas giant.

In most theories, $M_0$ depends on the distance $a_f$ of the growing 
protoplanet from the host star \citep{raf2006,raf2011,piso2014}. 
With no observational knowledge of $a_f$, we adopt two approaches.  
Our nominal calculations use $M_0$ = 10~\mearth. To allow for a 
plausible range of $M_0$, we also derive mass distributions when 
$M_0$ is randomly selected from a range of core masses:
\begin{equation}
\Ms = \left\{
\begin{array}{lll}
M_p & \hspace{10mm} & M_p \le M_{0,min} \\
r(M_{0,min},M_p) & & M_{0,min} < M_p < M_{0,max} \\
M_{0,max} & & M_p \ge M_{0,max} \\
\end{array}
\right.
\label{eq: m-solid2}
\end{equation}
where $M_{0,min}$ = 7.5~\mearth, $M_{0,max}$ = 12.5~\mearth,
and $r(a,b)$ is a random number uniformly distributed on the 
mass interval $(a,b)$.  This second approach follows the 
spirit of detailed calculations for the core mass without 
knowing $a_f$ for each star in the \kep\ sample.
Our conclusions are relatively insensitive to the values 
of $M_{0,min}$ and $M_{0,max}$ (\S 5.1).

This algorithm agrees reasonably well with observed estimates of 
heavy element abundances for exoplanets \citep[e.g.,][]{torres2008,
miller2011}. In small samples with state-of-the-art analyses, 
exoplanets have masses in heavy elements of 10--100~\mearth. 
Our adopted core masses lie at the lower limit of this range.
Exoplanets are enhanced in heavy elements relative to the host 
star; the enhancement is correlated with the stellar metallicity.
Without robust metallicity estimates for each \kep\ target, we
ignore this correlation.

To compare our \kep\ results with the \citet{mayor2011} data, we derive 
the differential incidence rate for the \kep\ transiting planets 
from \citet{batalha2013} with no correction for the likely gas 
fraction among planets with $\Mp > M_0$. Using the same mass bins
and an incidence rate of 54\% for 1.5--30~\mearth\ planets, our 
derived incidence rates for planets with $P \le$ 125~d 
(Figure~\ref{fig: exo-mass1}; solid blue curve) closely follow 
the HARPS results (Figure \ref{fig: exo-mass1}; orange curve). 
For $\Mp \approx$ 1.5--4~\mearth\ and 10--25~\mearth, 
the incidence rate of planets from HARPS is larger than the 
\kep\ rate. At 4--10~\mearth, the \kep\ incidence rate is somewhat
larger than the HARPS rate \citep[see also][]{dong2013}.

The dotted and dashed blue curves in Figure \ref{fig: exo-mass1} 
show the incidence rates for solids after correction for the 
likely gas fraction in massive planets. 
When $M_0$ = 10~\mearth, all planets with
$\Mp >$ 10~\mearth, have exactly 10~\mearth\ of solids. Thus,
the incidence rate for 6--10~\mearth\ planets jumps by more 
than 15\%; rates for more massive planets fall to zero 
(dotted blue curve).  Adopting a range of core masses, 
$M_0$ = 7.5--12.5~\mearth, leaves some planets in the 
10--15~\mearth\ bin (dashed blue curve). 

Anticipating the functional form of the mass distributions derived 
for the discs of T Tauri stars, we derive a cumulative mass 
distribution $f(>M)$ for the ensemble of \kep\ planet candidates 
from \citet{batalha2013}.  After estimating $\Ms$ for \kep\ planets 
with $\Rp \ge$ 1~\rearth\ and $P \le$ 125~d, the total mass for each 
system follows from adding the $\Ms$ for each planet in the system.
This estimate accounts for the multiplicity of each \kep\ planetary 
system.  After sorting the systems by total mass, we calculate the 
fraction of systems with total mass $f(> M)$ and normalize the total 
fraction for all systems to values appropriate for planets with 
$P\le$100\,d (0.515) and $P \le 400$~d \citep[0.77;][]{dong2013}.

Figure~\ref{fig: exo-mass2} shows the cumulative incidence rate 
for the total mass of solids in planetary systems from \kep.
This result assumes a random distribution of core masses, 
7.5--12.5~\mearth, and a total incidence rate of 0.77 for 
planets with $P \le$ 400~d. Because most of the uncertainty in the
cumulative rate lies at the smallest masses, the rate increases
from zero for the most massive planets to the total incidence rate
for the least massive planets.  At the largest masses 
($\Ms >$ 10~\mearth), the curve rises slowly with decreasing mass.
After a steep rise at 3--10~\mearth\ (see also 
Figure~\ref{fig: exo-mass1}), the curve turns over at the smallest
masses where the \kep\ data are less complete 
\citep[see also][]{dong2013,foreman2014}.

Changes in the adopted value for the mass of solids in a giant planet
have a negligible impact on the incidence rate. 
Alternative mass-radius relations for exoplanets \citep[e.g.,][]{weiss2014}
tend to enhance the incidence rate of 2--4\,\mearth\ planets at the
expense of 1--2\,\mearth\ and 4--10\,\mearth\ planets; using 
these relations 
steepens the cumulative distribution in the 3--5~\mearth\ region 
and flattens the cumulative distribution at lower masses. 
Adopting a different range of host star effective temperatures also 
has a small impact on the derived incidence rates.  Because lower 
mass stars tend to have lower mass planets \citep{john2010}, adding 
stars with lower effective temperatures raises (lowers) the incidence 
rate for lower (higher) mass planets.

Analyzing the larger set of exoplanet candidates from 
\citet{burke2014} also leads to similar results. With two extra 
quarters of data, the \kep\ Q1--Q8 sample contains a larger
fraction of smaller (lower mass) planets. Assuming a fixed 
incidence rate of planets at 1~\mearth\ (0.515 for $P \le$ 100~d
and 0.7725 for $P \le$ 400~d), the Q1--Q8 sample thus yields a 
somewhat smaller (larger) incidence rate at large (small) masses. 
Without a detailed analysis of the \kep\ detections as in 
\citet{dong2013}, however, the cumulative Q1--Q8 incidence 
rate is uncertain.  Lacking a robust normalization for the Q1--Q8 
data, we adopt results from the \citet{batalha2013} sample.

Given the good correspondence of the uncorrected \kep\ (solid
blue curves) and HARPS (orange curves) incidence rates in
Figure~\ref{fig: exo-mass1}, the \kep\ rates with corrections 
for the gas mass fraction (Figure~\ref{fig: exo-mass1}, dotted 
and dashed blue curves) provide a reasonable estimate for the 
incidence rate for solids in known short-period exoplanets. 
Our approach explicitly includes the observed multiplicity
of planets among {\it Kepler} host stars.
In the rest of the paper, we use the \kep\ results with a 
7.5--12.5~\mearth\ range in core masses to represent the population 
of planets with $P \le$ 400~d (Figure~\ref{fig: exo-mass2}).

To add information on the frequency of super-Earths and Neptunes
at larger $a$, we include results from comprehensive microlensing
surveys \citep[e.g.,][]{gould2010,cassan2012}.
Although the incidence rate of planets with $P \approx$ 100--400~d 
is approximately constant with log $P$,  
microlensing planet searches suggest a rapid increase in the incidence 
rate at much larger orbital periods. For orbital distances 
$a \approx$ 0.5--10~AU, the derived planet incidence rates 
from microlensing, 52$^{+22}_{-19}$\% for 10--30\,$\Mearth$ 
planets and 62$^{+35}_{-37}$\% for 5--10\,$\Mearth$ planets, are 
much larger than incidence rates at smaller $a$ \citep{cassan2012}. 
Figure~2 compares the 
incidence rate of the 5--10~\mearth\ and the 10--30~\mearth\ 
microlensing populations (two lower mass open magenta boxes) with 
the \kep\ P $\le$ 400~d cumulative incidence rate (green curve).  
The width of each open box represents the mass range associated 
with each incidence rate.  
The solid violet squares indicate the solid masses that are 
plausibly associated with these populations. 
Because published microlensing analyses select for binary events 
consisting of a star and a planetary-mass companion, the microlensing 
incidence rates apply to systems without a stellar companion within 100\,AU. 

\subsection{\bf Giant Planets:}

While only 2--3\% of stars host a giant planet within 100\,d
(Cumming et al.\ 2008; Mayor et al.\ 2011; Dong \& Zhu 2013),
the incidence rate of giant planets grows with orbital separation. 
Radial velocity surveys directly probe the frequency of gas giant
planets for orbital periods $P$ = 2--2000~d \citep{cum2008,mayor2011}.
Extrapolating the $d\log M/d\log P$ distribution to longer periods 
suggests a giant planet incidence rate of 17\%--20\% within 20\,AU 
(Cumming et al.\ 2008).

Microlensing studies also derive large incidence rates for giant planets.
Among K--M dwarfs in the Galaxy, 17$^{+6}_{-9}$\% have a Jupiter-mass 
planet (0.3--10\,$M_J$) with $a \approx$ 0.5--10\,AU \citep{cassan2012}. 
Over a similar range of masses and semimajor axes, \citet{gould2010} 
derive a frequency of 36$\pm$15\%.  
Although this orbital period range is somewhat larger than the range 
probed directly by radial velocity surveys, the extrapolated rate from 
\citet{cum2008} agrees remarkably well with the microlensing rates
\citep[see also][]{gould2010,clanton2014}.

Figure~\ref{fig: exo-mass2} compares the incidence rates for 
gas giants from radial velocities
(Cumming et al.\ 2008; open deep blue box) and from
microlensing (Cassan et al.\ 2012; highest mass open magenta box) 
with the \kep\ P $\le$ 400~d rates (green curve). 
The width of each open box represents the mass range associated with 
each incidence rate.
To inventory only the solids associated with these populations, 
we adopt a solid core mass of 
10~\mearth\ 
for both the radial velocity and microlensing giant planet populations. 
The incidence rates are shown as filled squares centered at this value.

As shown in Figure~\ref{fig: exo-mass2}, the two independent estimates 
of the giant planet incidence rate at 1--20\,AU 
(from radial velocities, solid deep blue box; and 
microlensing, lowest solid violet box) 
agree to $\pm$ 2\%. 
These rates are only a few percent larger 
than the frequency of $>10$~\mearth\ \kep\ planets with $a \le$ 1~AU 
(green curve).

For 5--30~\mearth\ planets,  the differential incidence rate from
microlensing 
(upper two solid violet squares) 
is clearly much larger 
than the HARPS or \kep\ rates for planets at P~$\le$400~d. Both of
these rates are also much larger than the rates for gas giants.
Overall, these two microlensing samples represent the largest 
reservoirs of super-Earth and Neptune mass planets.

\subsection {\bf Debris Discs }

With typical fractional luminosities of $L_d/L_{\star} \approx 10^{-5}$ to 
$10^{-3}$, the IR excesses of debris disc systems 
require as much as a few lunar masses in small 
(1~$\mu$m to 1~mm) dust grains \citep{bac1993,wyatt2008,matthews2014}. 
Typical temperatures of 50~K to 300~K imply rings or discs of dust at 
distances of a few AU to a few hundred AU from the host star 
\citep{wyatt2008,chen2014}.
Because radiation pressure and destructive collisions remove small 
particles on short time scales, these high frequencies require an 
invisible supply of material (i.e., larger ``parent bodies'') to 
replenish the dust. Planetesimals with radii of 1--100~km are an 
obvious choice \citep{aum1984}.  
In this picture, a cascade of collisions among 1--100~km and smaller objects 
produces the 1--1000~$\mu$m particles that emit the IR excess. As the cascade 
proceeds, radiation pressure ejects 1--10~$\mu$m particles from the disc. Over time, 
the cascade removes nearly all of the material originally present in the 
1--100~km parent bodies.  

The commonality of debris discs suggests that many stars harbor large 
reservoirs of solids at distances beyond 20--30~AU 
\citep{wyatt2008,kb2008,kriv2008,kb2010,kw2010}. 
Extensive surveys with {\it IRAS}, {\it ISO}, and {\it Spitzer} find 
debris discs around 10\% to 20\% of solar-type main sequence stars 
\citep{bry2006,trill2008,carp2009,sier2014}. 
Deeper surveys with {\it Herschel} also identify debris discs associated 
with roughly 20\% of solar-type main sequence stars \citep{eiroa2013}. 

Recent data suggest that the frequency of debris discs is fairly independent of local 
environment.  Among all FGK stars, single and binary stars are equally likely to 
have debris discs \citep{trill2007}.  For FGK stars with known planets, the incidence 
rate of $\sim$ 29\% is similar to the rate of $\sim$ 20\% for all FGK stars 
\citep[][see also Bryden et al. 2009]{eiroa2013,marshall2014}.  Among the 
planet-bearing population, debris discs are more common in systems with low mass 
planets (6/11 systems with $\Mp \le 30~\Mearth$) than with higher mass planets 
(5/26 systems with $\Mp > 30~\Mearth$).

From analytic and numerical calculations, explaining the frequency and 
level of debris disc emission requires 10--100~\mearth\ of solids in 
parent bodies 
beyond 10--30~AU \citep{hab2001,wyatt2008,kb2008,kb2010,kw2010,gaspar2013}.
Thus, the current picture of debris disc evolution eliminates 
the 10--100~\mearth\ stored in parent bodies from the solid 
reservoir available to make the known exoplanets 
\citep{wyatt2008,kb2008,kriv2008,raymond2011}.

To place the mass distribution of debris discs on the same footing as exoplanets,
we adopt a simple prescription. 
We assume the debris disc incidence rate from {\it Herschel} of 
$\sim 20$\% among solar-type stars.  
We also assume that a debris disc indicates at least 10~\mearth\ in 
1--100~km objects.
Explaining observations
of debris discs therefore requires that 20\% of solar-type stars have at least 
10~\mearth\ in solids which does not participate in the formation of Earth-mass 
and larger planets. For a first comparison, we assume that this 20\% is distributed among
systems with 10--20~\mearth\ in solids (Fig.~2; orange box). 
The placement of the orange box
suggests that debris discs require a large amount of mass, comparable to the mass 
required for gas giants and short period exoplanets from HARPS and \kep.

Overall, the fraction of stars with debris discs beyond 20--30~AU is comparable to 
the fraction of stars with gas giant planets within 20~AU (Fig.~2; 
orange and blue boxes). The frequency of both groups is also similar to the fraction 
of 10~\mearth\ or larger planets within roughly 1~AU. Each of these fractions is much 
smaller than the fraction of 5--30~\mearth\ planets detected with microlensing.

\section{Reservoirs of solids in protoplanetary discs}
\label{sec: discs}

For more than fifty years, the Taurus-Auriga molecular cloud has been 
the gold standard of low mass pre-main sequence stellar evolution 
\citep[see, for example,][and references therein]{cohen1979,kh1995,kgw2008,luhman2010}. 
With roughly 350 fairly isolated young stars that span 
a range of 
evolutionary stages, Taurus-Auriga has become a popular laboratory 
to investigate the properties of young stars and their circumstellar
discs. Although other star-forming regions (e.g., IC 348, $\rho$ Oph,
and the Sco--Cen association) also provide vital information on
the evolution of pre-main sequence stars and their discs
\citep[e.g.,][]{and2007,herbst2008,preib2008,wilk2008,will2011}, 
the physical properties of the discs in these regions are fairly similar 
to the larger ensemble of discs in Taurus-Auriga \citep[e.g.,][]{and2013}.

In our analysis, we rely on previous classifications of the evolutionary
state of the central pre-main sequence star in each system. From analyses 
of spectral energy distributions (SEDs) \citep[e.g.,][]{adams1987}, stars 
are divided into class~I (protostar, a central star surrounding by a disc
and an infalling envelope),
class~II (classical T Tauri star, a central star surrounded by an opaque 
disc, with little or no envelope), and
class~III (weak emission T Tauri star, a central star with little or no 
disc and envelope). 
Class~I protostars constitute roughly 10\% of the complete Taurus-Auriga 
sample and have typical lifetimes of $\sim$ 0.3 Myr \citep{offner2011};
class~III sources comprise roughly 25\% of the complete class~II + III 
population for solar-type stars \citep{luhman2010}. 
The discs in class~II sources are commonly assumed to be close analogues 
of the discs from which the solar system and the known exoplanets formed
\citep[e.g.,][]{hues2005,ida2005,bk2011a,raymond2011}. 

Submillimeter continuum observations constrain the mass of solids in 
nearby circumstellar discs.  \citet{and2013} report disc masses for 
an extensive sample of Class~II stars in Taurus-Auriga. Based on the
\citet{luhman2010} study, this sample is statistically complete for 
stellar spectral types earlier than M8.5. Although \citet{and2013}
exclude class~I and class~III sources from their analysis, 
\citet{and2005} summarize disc masses for a less complete set of 
class I, II, and III objects.

The Andrews et al.\ (2013) 
disc masses $M_d$ 
assume 
a simple relation between disc mass and submillimeter flux:
\begin{equation}
M_d = {d^2 F_\nu \over \kappa_\nu B_\nu(T_c)}
\end{equation}
where $F_\nu$ is the submillimeter flux density, $d$ is the distance, 
$\kappa_\nu$ is the opacity, and $B_\nu(T_c)$ is the Planck function 
at a characteristic temperature $T_c$ that depends on the stellar luminosity.
The main uncertainties in disc masses result from 
uncertainties in $T_c$ 
and the dust 
opacity \citep[see][]{and2013}, 
the latter arising in part from uncertainties in the 
composition and compactness of the solids.  
Although these uncertainties have little 
impact on our analysis, we return to them in 
\S\ref{sec: discuss}.

The resulting cumulative distribution of Class~II disc masses as a fraction 
of the stellar mass $f(>M_d/M_*)$ rises approximately linearly with decreasing 
$\log M_d$, from an upper mass limit of $M_d\sim 0.05\,M_*$, down 
to a lower mass limit of $ M_d\sim 0.00015\,M_*$ where it reaches unity
(Fig.~\ref{fig: disc-mass}, grey curves 
for a stellar mass of $1~\Msun$).
Because Andrews et al.\ (2013) assume a gas-to-dust ratio of 100
\citep{and2005}, these upper and lower limits correspond to disc 
solid masses of $\Ms = 167\,\Mearth$ and $0.5\,\Mearth$ for 
a $1\,\Msun$ star.
The median disc in the sample
has a mass of
$M_d = 0.003 ~ \Msun$ in gas
and 
$\Ms = 10 ~ \Mearth$ in solids.
Some 70\% of discs have masses larger than
$M_d > 0.0008 ~ \Msun$ or
$\Ms > 2.7 ~ \Mearth$.

The class~II disc mass distribution 
includes both single, binary, and multiple star systems.  In 
contrast, the HARPS and microlensing studies explicitly exclude stellar 
binaries \citep{mayor2011,cassan2012}.  Because discs in binaries are 
typically less massive than discs in single stars 
\citep{jensen2003,and2005}, 
including multiples in the submillimeter sample skews the disc 
mass distribution to lower masses.
However, the impact of a stellar companion on disc mass is a strong 
function of the binary separation.  In Taurus-Auriga, binaries with 
separations $> 300$\,AU have 880 $\mu$m continuum fluxes very similar 
to those of single stars \citep{harris2012}. For closer binaries, 
the fluxes range from 5 times smaller for separations of 30--300~AU 
to 25 times fainter for separations $<$ 30~AU. These results indicate 
that an appropriate sample of `single' stars excludes binaries with 
separations less than 300~AU.

To construct a sample of class~II T Tauri stars free of close binaries, 
we remove known
binaries with separations of $\le 300$~AU \citep{kgw2008,luhman2010,kraus2011}
from the \citet{and2013} sample.  This conservative cut reduces the complete 
Andrews et al.\ (2013) sample from 210 stars to 152 stars. As in \citet{and2013}, 
we use the Kaplan-Meier estimator to derive the mass distribution for a set
of sources with clear detections and upper limits.  The resulting distribution 
of Class~II single star disc masses $f_s(>M_d/M_*)$ 
(Figure~\ref{fig: disc-mass}, cyan curve) is slightly steeper than 
the \citet{and2013} distribution (grey curves). 
The median mass of $\sim 13\,\Mearth$ is roughly 50\% larger than the median 
disc mass for the complete \citet{and2013} sample. Despite the difference in
median mass, the masses for discs in the top 1\% and 10\% and the bottom 1\% 
and 10\% are indistinguishable.

Although the Taurus-Auriga class~II sample is nearly complete for spectral 
types earlier than M8--M9, binary samples are relatively incomplete for 
stars with spectral types later than roughly M4 \citep{kraus2011}. To
investigate whether $f_s(>M_d/M_*)$ is sensitive to the lower spectral type 
limit, we created a sample of 95 stars consisting only of apparently single 
class~II stars with spectral types of M4 and earlier. The resulting mass 
distribution is indistinguishable from the distribution for all spectral types.  

As a final attempt to characterize how the mass distribution depends on the
evolutionary state of a pre-main sequence star, we consider the impact of 
the `discless' class~III sources in Taurus-Auriga.  Submillimeter surveys 
of these weak emission T Tauri stars are not as complete as those for the
classical T Tauri stars 
\citep{and2005,and2013}. 
However, we can make a simple estimate of 
the impact of these 
objects on the mass distribution. 
Assuming that the class~III population (which represents 25\% of the combined 
class~II + III population) is discless, 
we add an ensemble of class~III sources 
with zero disc mass to the complete set of class~II sources from \citet{and2013}. 
The resulting mass distribution is shallower 
(Figure \ref{fig: disc-mass}, red curve); the median mass drops to roughly 
7~\mearth.

The combined class~II + class~III mass distribution, shown as the red curve in 
Figure \ref{fig: disc-mass}, represents a rough lower limit to the mass reservoir
available for planet formation. To set a rough upper limit to this reservoir,
the violet curve plots the mass distribution for the Taurus-Auriga class I sources 
\citep{and2005}. This distribution is much steeper than the class~II or 
class~II + III distributions. The median mass reservoir is 50--100~\mearth; 
90\% of the sources have reservoirs of 20--1000~\mearth.  With a median mass of 
roughly 10--15~\mearth, the discs in the complete ensemble of class~II sources 
(Figure \ref{fig: disc-mass}, blue or grey curves) are 5 times less massive than 
the class I discs.

\section{Comparison of Solids in Planetary Systems and Class~II Protoplanetary Discs}
\label{sec: comp}

\begin{figure}
\includegraphics[width=17.5cm]{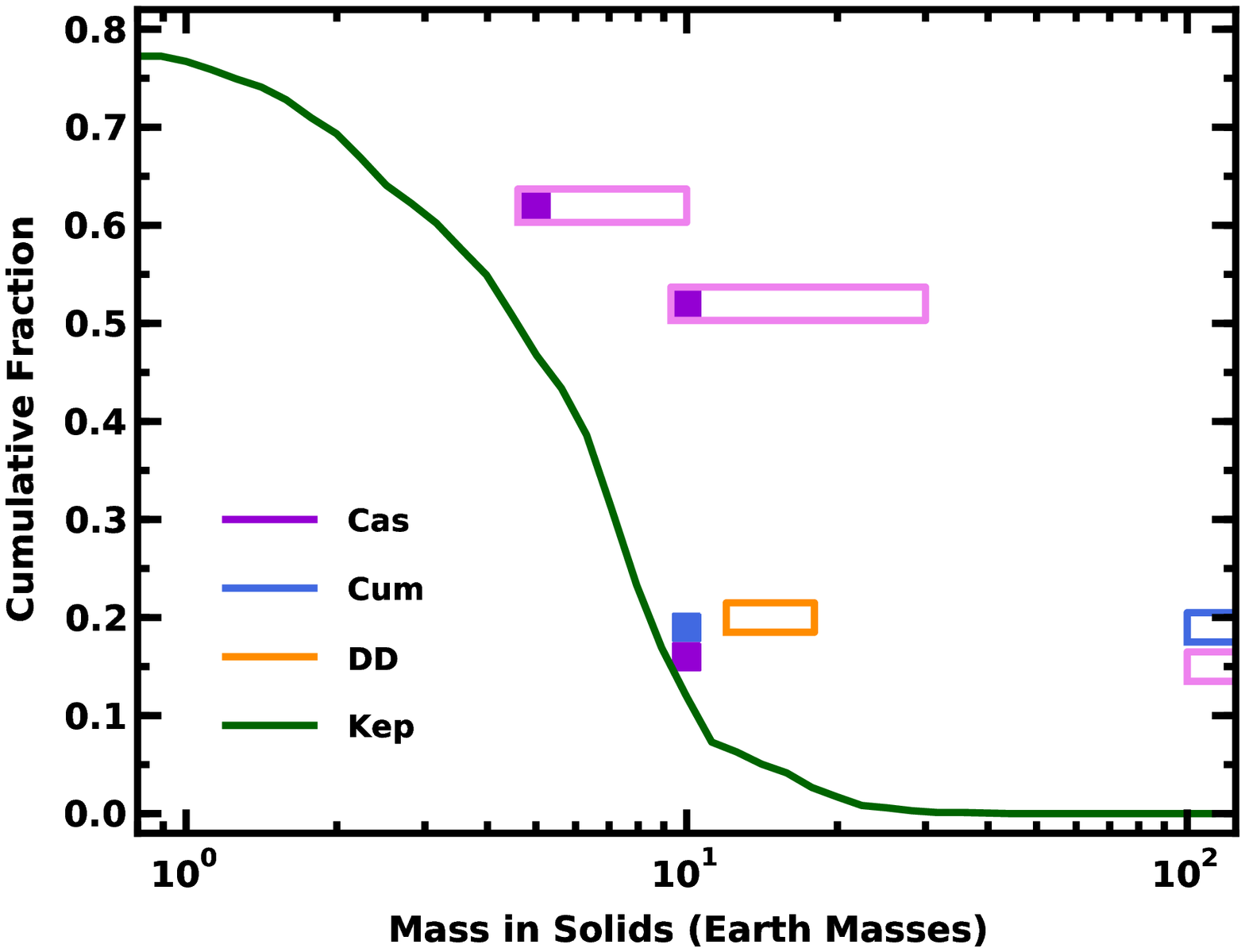}
\caption{
Differential incidence rates for 
debris discs \citep[`DD'; e.g.,][]{eiroa2013} and
exoplanets from
\kep\ (`Kep', this paper),
microlensing \citep[`Cas';][]{cassan2012}, and
radial velocities \citep[`Cum';][]{cum2008}.
Solid green curve: cumulative incidence rate for 
\kep\ planets with $P \le$ 400~d.
Violet boxes: differential rates for microlensing planets
with $a$ = 0.5--10~AU. 
Open, light rectangles indicate the nominal mass range;
dark, filled squares show the adopted mass in solids.
Blue boxes: nominal mass range (open symbol) and adopted 
mass in solids (filled symbol) for gas giant planets with 
$a \le$ 20~AU inferred from radial velocity surveys.
Orange rectangle: adopted range in solid mass at $a \ge$ 20~AU 
for debris discs.
\label{fig: exo-mass2}
}
\end{figure}

\begin{figure}
\centering
\includegraphics[width=17.5cm]{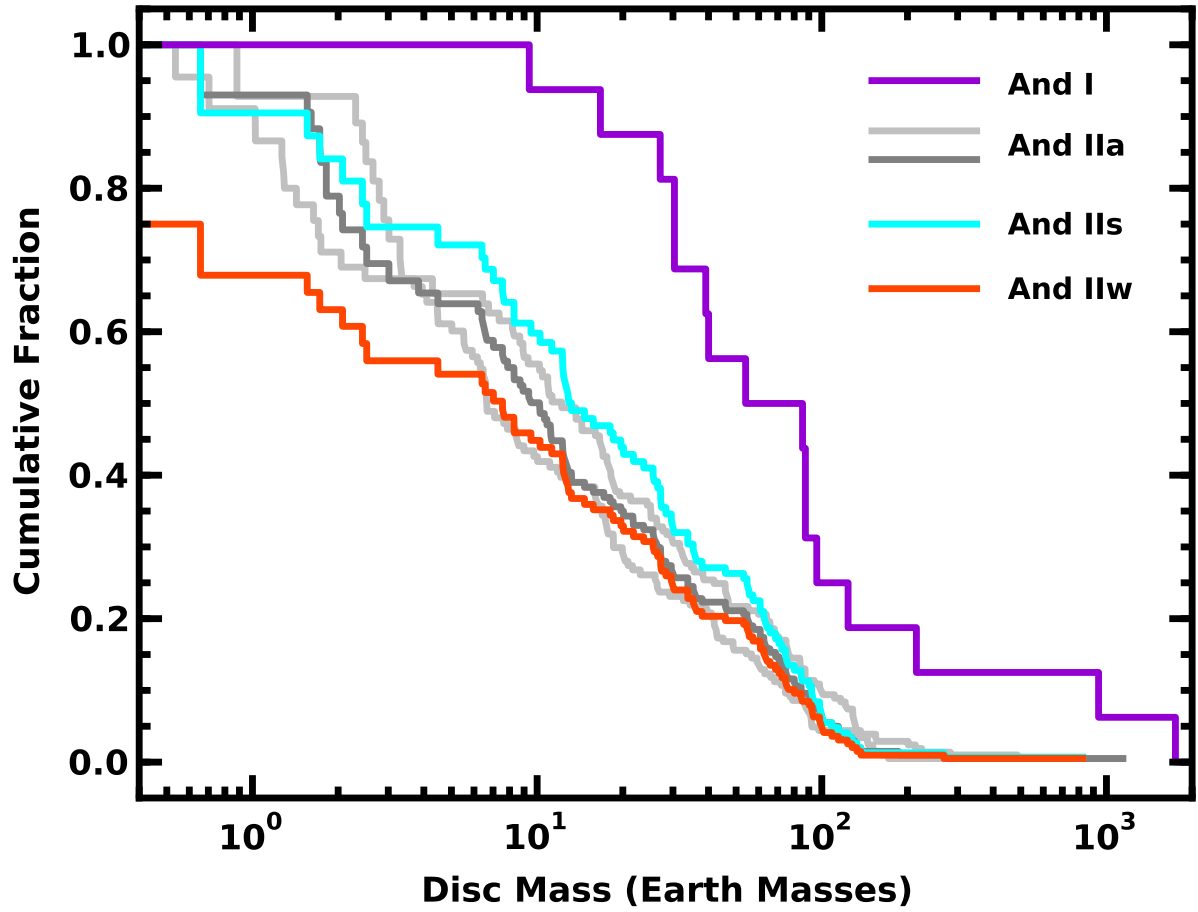}
\caption{
Cumulative mass distributions ($f (> M_d / M_*)$) for protoplanetary discs. 
Violet curve (`And I'): class I sources from \citet{and2005}
Grey curves (`And IIa'): complete set of class~II sources from 
\citet{and2013} using three different calculations of the 
mass-luminosity-temperature relation for pre-main sequence stars 
\citep[for details, see][]{and2013}.
Cyan curve (`And IIs'): set of `single' class~II sources with the 
mass-luminosity-temperature relations from \citet{and2013}
and the \citet{siess2000} evolutionary tracks.
Red curve (`And IIw'): class~II +~III sources assuming a 25\% fraction of
discless stars.
\label{fig: disc-mass}
}
\end{figure}

From a comparison of Figures 2 and 3, it is clearly difficult for the class II 
disc mass budget to account for the solids bound up in the known exoplanet 
and debris disc populations. 
With reservoirs of 10~\mearth\ required for 
each of debris discs (Figure~\ref{fig: exo-mass2}, orange box), 
Neptunes in microlensing surveys (Figure~\ref{fig: exo-mass2}, middle purple box), 
and gas giants (Figure~\ref{fig: exo-mass2}, deep blue box), 
and with their respective incidence rates of $\sim 20$\%, $\sim 50$\%, and 
$\sim 20$\%,  it seems 
clear that any class~II + III sample with a median mass of 7~\mearth\ is 
hard-pressed to supply enough material to form even this subset of known 
exoplanetary systems. 
Even at the higher median mass of the single star class~II sample (10--15~\mearth), 
these three populations alone 
are a severe
burden on the class~II mass budget.  To consider these conclusions in more detail, 
we now develop several quantitative comparisons between the solid masses available in 
protoplanetary discs and the solid masses required for known planetary systems.

In \S 4.1, we use a simple tally approach that considers the solid masses
available in the outer ($>20$\,AU), inner ($< 4$\,AU), and middle ($4-20$\,AU)
regions of the disc, compared with the demands placed on these solid
reservoirs by debris discs, Kepler planets, and the remaining
planet populations, respectively. In \S 4.2, we use a Monte
Carlo approach to create ensembles of systems with planets and
debris discs at their known incidence rates; we then compare the 
solid mass distribution of the ensembles with that of protoplanetary
discs. Compared to a simple tally, the Monte Carlo analysis considers 
only the total solid mass of each system and ignores the fractional 
disc mass at different radii that is available to form the planet and
debris disc populations.

\subsection{A Simple Tally}
\label{sec: tally}

Only a fraction of the total disc mass is likely to be available 
to generate specific populations of planets and debris discs. 
For example, the 
(i) \kep\ planets ($\Mp$ = 1--30~\mearth, $a \le$ 1~AU, $P \le$ 400~d),
(ii) microlensing planets ($\Mp > 5~\mearth$, 
$a \approx$ 0.5--10~AU, $P \approx$ 0.35--30~yr), 
and 
(iii) debris discs ($a \ge$ 20~AU, $P \ge$ 100~yr) 
are expected to have formed from solids in the inner, middle, 
and outer regions of the disc. 
Therefore, a simple way to compare the solid distributions in 
planetary systems and protoplanetary discs is to compare the planet 
mass distribution in 3 radial bins with the disc mass distribution 
in similar radial bins.

For this comparison, 
we first separate the
single-star class~II disc mass reservoir into an outer disc ($>20$\,AU)
that may generate debris and an inner disc ($\le20$\,AU) that produces 
the known planets.
This division 
is motivated by the typical semimajor axis ranges of 
(i) the parent bodies of debris discs (typically $a \ge$ 20~AU) 
and
(ii) the \kep\ and microlensing planets ($a \le$ 20~AU). 

To construct these reservoirs, we assume that class~II discs have 
power-law surface density distributions $\Sigma \propto a^{-p}$,
with $p$ = 1.5 as in the MMSN, and outer radii, 
$r_{out}$ = 50~AU. Compared to a flatter $p=1$ distribution, 
which is often invoked as characteristic of a 
steady accretion disc, a surface density 
distribution as steep as $p=1.5$ concentrates more of the solids at 
small radii, an arrangement that helps to account for the large solid 
mass that is bound up in the $P<400$\,d planet population 
(see also Chiang \& Laughlin 2013).

\begin{figure}
\centering
\includegraphics[width=15cm]{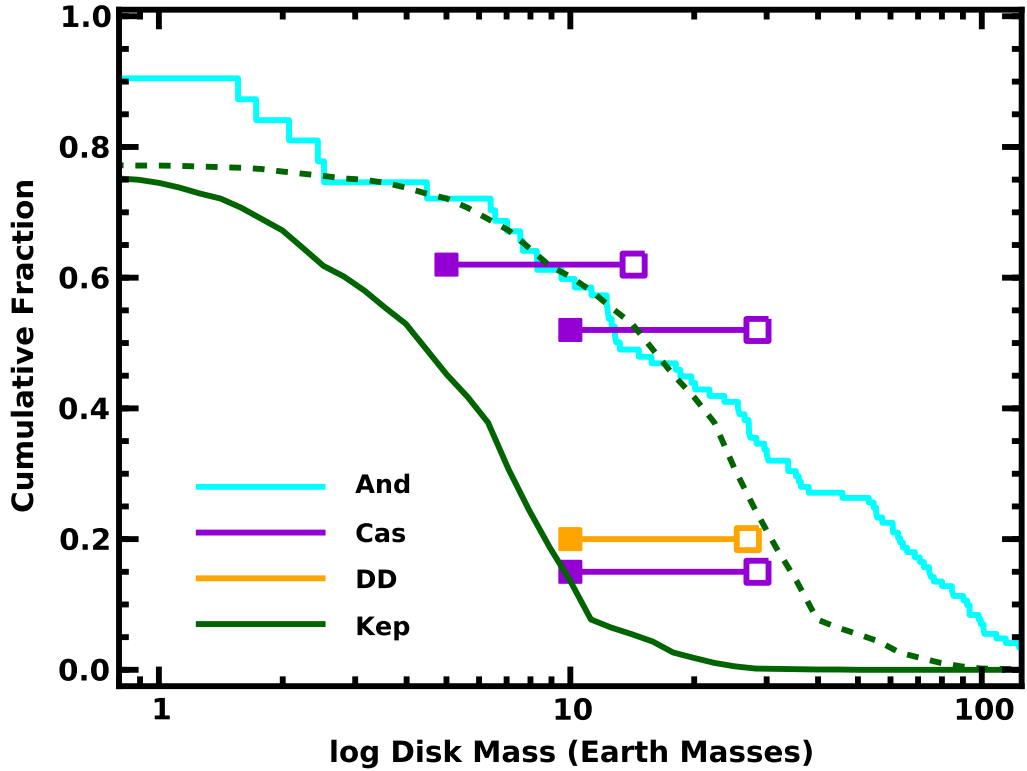}
\caption{
Mass distributions for protoplanetary discs.
Cyan curve (`And'): observed mass distribution for 
single class~II sources
from Figure~\ref{fig: disc-mass} \citep{and2013}.
Solid curve and filled symbols: observed mass distribution for \kep\ planets 
with $P \le$ 400~d (`Kep'; green curve'), 
debris discs (`DD'; orange box), and 
microlensing planets (`Cas'; violet boxes).
Dashed curve and open symbols: 
required total disc masses $M_{tot,s}$ if $\Sigma \propto a^{-3/2}$ and 
if the above planets and debris discs arise from 
(i) $a \le$ 4~AU (dashed green curve, Kepler planets with $a \le$ 1 AU),
(ii) 4~AU $\le a \le$ 20~AU (open violet boxes, microlensing planets), and
(iii) $a \ge$ 20~AU (open orange box, debris discs).
\label{fig: exo-disc1}
}
\end{figure}

If the observed dust emission from debris discs persists over the main 
sequence lifetime of the host star, analytic and numerical models 
require solid reservoirs of at least 10~\mearth\ beyond 20\,AU 
\citep[Figure~\ref{fig: exo-disc1}, solid orange box; see][and 
references therein]{wyatt2008,kb2008,kb2010,kw2010}.  
If $\Sigma \propto a^{-1.5}$ and $r_{out}$ = 50~AU, the outer disc 
(20--50\,AU) contains 37\% of the total mass in solids.  Thus, debris disc 
systems are then 
drawn from discs with total solid masses $M_{tot,s}$ at least 
1/0.37 times larger than 10~\mearth, $M_{tot,s} \ge 27\,\Mearth$
(Figure~\ref{fig: exo-disc1}; open orange box). 
The fraction of discs in the single-star class~II distribution 
with
masses in this range ($\sim 40$\%; cyan curve) can easily explain 
the debris disc incidence rate ($\sim 20$\%; open blue box).

With the outer 37\% of the disc mass set aside for debris discs, we now divide
the remaining 63\% into separate reservoirs for the \kep\ planets 
and the microlensing planets. The similarity in the shape of the solid green 
and cyan curves in Figure~\ref{fig: exo-disc1} suggests that we can reduce 
the masses of single, class~II discs by a factor of 3.5 to achieve a 
rough match with the incidence rate for \kep\ planets. For a disc
with $\Sigma \propto a^{-3/2}$ and $r_{out}$ = 50~AU, 28\% (1/3.5) 
of the mass is contained within $a \le$ 4~AU. Therefore, if {\it every}
\kep\ planet is drawn from a disc roughly 3.5 times more massive 
than the planet
(Figure~\ref{fig: exo-disc1}, dashed green curve), 
the mass contained in {\it all} \kep\ planets
will not overly tax 
the solid reservoirs in class~II discs (Figure~\ref{fig: exo-disc1}, 
cyan curve).  Producing the $P\le400$~d \kep\ planets from solids 
within 4~AU requires some mechanism to concentrate the solids toward 
smaller disc radii.  We return to this point in \S5.

Having reserved the inner 28\% of the solids ($a \le$ 4\,AU) for the
formation of the $P\le400$\,d planet population and the outer 37\% of the
solids ($a \ge$ 20~AU) for the generation of debris discs, we 
are left with the 35\% of solids in the middle region (4--20\,AU) of 
the disc.  This reservoir needs to have enough mass to account for 
the remaining planet populations 
-- super-Earths (5--10\,$\Mearth$), Neptunes (10--30\,$\Mearth$),
and gas giants ($>100\,\Mearth$) -- located at 0.5--20\,AU.
To evaluate the ability of this middle disc region to produce
these planets, we assume that 
super-Earths have 5~\mearth\ of solids;
Neptunes and gas giants each have 10~\mearth\ of solids
(Figure~\ref{fig: exo-disc1}, filled purple boxes).
If these planets arise from the middle region of the disc, they 
are drawn from discs with masses 1/0.35 larger,
$M_{tot,s} \ge 14\,\Mearth$ for super-Earths, 
and $M_{tot,s} \ge 28\,\Mearth$ for Neptunes and 
gas giants (Figure~\ref{fig: exo-disc1}, open purple boxes).

The middle disc reservoir can easily manage to explain the 
frequency of gas giants at 0.5--20~AU. The solids in gas 
giants require that roughly 17\% of discs have masses of 
$M_{tot,s}\ge 28\,\mearth$.
Roughly 40\% to 50\% of class~II discs meet this criterion
(Figure~\ref{fig: exo-disc1}, cyan curve).

Despite this success, it is impossible for the middle disc
reservoir to explain the high incidence rates of super-Earths 
and Neptunes at 1--10~AU. The super-Earths from microlensing
surveys require that 62\% of discs have masses of 
$M_{tot,s} \ge
14~\mearth$. Fewer than 50\% of discs match this constraint.
When combined with the 17\% incidence rate for gas giants
at 1--10~AU, matching the 52\% incidence rate of Neptunes 
is even more challenging. Roughly 40\% of class~II discs 
have 
$M_{tot,s}\ge 28\,\mearth$, well below the nearly
70\% required to explain the incidence rate of 
exoplanets with $\Ms > 10\,\mearth$ 
(Neptunes and gas giants) at 0.5--10~AU.

Modifying the sizes of the three disc reservoirs cannot change
these conclusions.  For example, the incidence rate of 
5--30~\mearth\ microlensing planets requires that {\it every disc} 
have at least 5\,\mearth\ in the middle disc region. 
However, only $\sim 80$\% of class~II sources have this much 
solid mass in the {\it entire disc}. 
Increasing the total mass in the middle disc reservoir by a factor
of 2--3 would yield solid masses sufficient to explain the microlensing
planets. However, removing this mass from the inner and outer
reservoirs leaves little or no (or negative) mass to match the
incidence rates for Kepler planets or debris discs.
Thus, the single star class~II disc mass distribution is insufficient 
to produce the known exoplanets.
The gap between these two solid reservoirs 
is probably broadened by additional factors 
(e.g., the efficiency of planet formation and 
undiscovered planet populations; see \S 5).

\subsection{Monte Carlo Simulation}
\label{sec: montecarlo}

To illustrate the challenges in more detail, 
we can compare the total solid masses bound up in 
planets and debris discs 
with those present in class II discs, independent of how the solids are 
distributed in radius. 
For this comparison, we create synthetic distributions 
of solids in planets and debris discs. 
Constructing ensembles of planets allows us to make a more 
general comparison with the mass distribution of T Tauri discs 
and to evaluate the sensitivity of the results to our
set of input assumptions.

To build a simple Monte Carlo simulation, the
basic input parameters are 
the fractions $f_j$ describing the incidence rate for each planet 
population: 
the fraction of stars with a debris disc 
($f_{dd}$), gas giant planet ($f_{gg}$), 
super-Earth (5--10~\mearth) microlensing planet ($f_{se}$), 
Neptune (10--30~\mearth) microlensing planet ($f_{nep}$), 
or a \kep\ planet ($f_k (\Ms)$). 
At the start of each simulation, all fractions are fixed, with four of the fractions --
$f_{dd}$, $f_{gg}$, $f_{se}$, and $f_{nep}$ -- having single values. \kep\ planets
are selected from a table derived from the \kep\ data described in \S2 with
a normalization factor $f_{k,0}$ which represents the total fraction of all
\kep\ stars with transiting planets having $P \le$ 400~d \citep{petig2013,dong2013}.

Each planet or debris disc has an assigned mass in solids $M_j$. For debris 
discs, gas giants, and microlensing planets, the constant core mass model 
has $M_{dd}$ = 10~\mearth, $M_{gg}$ = 10~\mearth, $M_{se}$ = 5~\mearth, and
$M_{nep}$ = 10~\mearth. \kep\ planets have masses assigned from eq.~(2). 
To investigate the sensitivity of our results to these parameters, we 
consider a variable core mass model where we assign a mass range for the 
core mass and select masses randomly distributed on the mass range as in
eq.~(3). In this set of models, 
$M_{dd}$ = 10--100~\mearth, $M_{gg}$ = 7.5--12.5~\mearth, $M_{se}$ = 
2.5-7.5~\mearth, and $M_{nep}$ = 7.5--12.5~\mearth. The cores of \kep\ 
planets have masses assigned from eq.~(3). 

To set the fractions, we adopt the observed values described in \S2.
For debris discs and gas giants, the fractions are straightforward,
$f_{dd}$ = 0.2 and $f_{gg}$ = 0.2. 
For the \kep\ planets, we adopt $f_{k,0}$ = 0.75 for transiting planets 
with orbital periods $P \le 400$~d \citep{dong2013}. Although the nominal 
fractions for the 
microlensing planets are $f_{se}$ = 0.62 and $f_{nep}$ = 0.52, the semimajor
axis range of the microlensing planets, 0.5--10~AU, overlaps with that of the
\kep\ planets ($a \le 1.05$ AU for $P \le$ 400~d around solar-type stars).  
Because the \kep\ incidence
rate has a smaller uncertainty, we prefer to use the longer period set of
\kep\ planets than the microlensing planets. To eliminate double-counting,
we then need to reduce the microlensing fractions. As guidance, 
\citet{cum2008} show that the frequency of giant planets slowly increases
with orbital period beyond $P =$ 300~d. Although the \kep\ data suggest
an incidence rate that is roughly constant with $\log P$ 
for $P$ = 100--400~d
\citep[e.g.,][]{petig2013,dong2013}, the variation of the incidence rate
with $P$ beyond 400~d is unknown. For simplicity, we assume
the incidence rate is independent of period, implying $f_{se}$ = 0.59 and 
$f_{nep}$ = 0.49 for $a$ = 1--10 AU.

To run the simulation, we draw random numbers $r_j$ ($j$ = 1--5) and
select the constant core mass model for the Kepler mass distribution. 
Four random numbers establish the 
presence or absence  
of a debris disc, a gas 
giant, or a microlensing planet:
\begin{equation}
p_j = \left\{
\begin{array}{lll}
1 & \hspace{10mm} & r_j \le f_j \\
0 & & r_j > f_j \\
\end{array}
\right.
\label{eq: prob}
\end{equation}
where $j$ = 1 (debris disc), $j$ = 2 (gas giant), 
$j$ = 3 (5~\mearth\ microlensing planet), and
$j$ = 4 (10 ~\mearth\ microlensing planet).
When $j$ = 5 (\kep\ planet) and 
$r_j \le f_{k,0}$, the 
random number sets the mass of the planet along the 
\kep\ mass distribution.

In the constant core mass algorithm, the total mass in planets for the 
$i$th star is then
\begin{equation}
M_{t,i} = M_k(p_5) + \sum_{j=1}^{j=4} ~ p_j ~ M_j ~ ,
\end{equation}
where the $M_j$'s are constant masses for debris discs, gas giants,
and microlensing planets.

In the variable core mass algorithm, we draw four additional random numbers
and select the variable core mass model for the total mass distribution of
\kep\ planets. The masses
$M_j$ ($j$ = 1--4) are then randomly selected on the appropriate range for
debris discs, gas giants, and microlensing planets.

After drawing sets of random numbers for $N$ stars, we sort the masses and
derive a cumulative mass distribution. Running $M$ trials of $N$ stars 
allows us to construct a median mass distribution and to estimate the 
inter-quartile range about this median. Tests with $N$ = 10000 and $M$ =
10000 demonstrate that the median and inter-quartile range are 
indistinguishable from the average and dispersion. The inter-quartile 
range and dispersion are always less than 1\%. Thus, the scatter about
the median mass distribution is negligible.

The multiplicity statistics of the simulated populations agree
well with observations. Roughly 75\% of the simulated systems have
multiple planets. Approximately 20\% of systems with a Kepler planet
have multiple Kepler planets \citep[cf.][]{rowe2014,fab2014}.  Among 
debris disc systems, $\sim 25$\% have a gas giant, and $\sim 75$\% have a 
\kep\ planet \citep[cf.][]{marshall2014}.

\begin{figure}
\centering
\includegraphics[width=15cm]{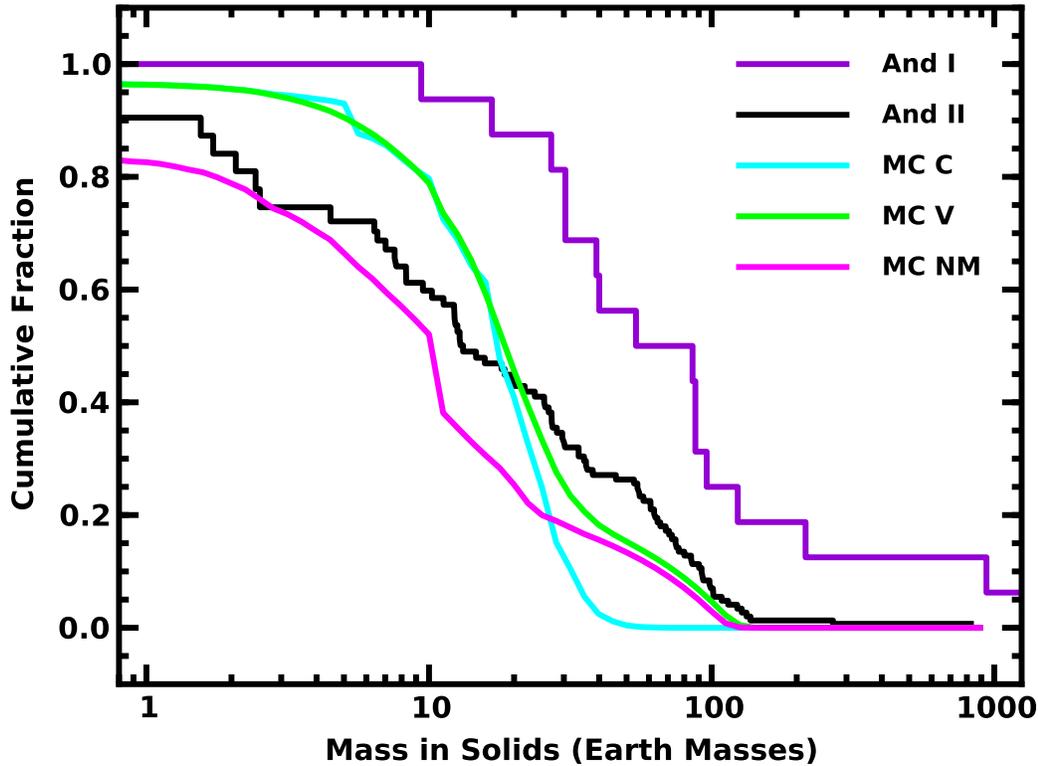}
\caption{
Comparison of incidence rates for exoplanets and protoplanetary discs.
Violet (`And I'; class I sources) and
black (`And II'; single class~II sources) jagged curves: 
observed mass distributions for protoplanetary discs \citep{and2013}.
Cyan (`MC C'; constant core mass model) and
green (`MC V'; variable core mass model) smooth curves:
mass distributions calculated from the Monte Carlo model
described in the text.
Smooth magenta curve (`MC NM'): mass distribution from the Monte Carlo
model with a variable core mass but no 5--10~\mearth\ planets 
at 1--10~AU from microlensing.
Although discs around T Tauri stars have enough mass to explain the
frequency of debris discs beyond 20~AU, gas giants inside 20~AU, and
\kep\ planets inside 1~AU, they do not contain enough mass to explain
the observed frequency of super-Earths detected in microlensing surveys.
Disks around protostars have large enough reservoirs to explain the
incidence rates for all known exoplanets.
\label{fig: exo-disc2}
}
\end{figure}

The Monte Carlo simulations confirm our conclusion that the masses of 
class~II discs are insufficient to explain the masses of all known 
exoplanets (Figure~\ref{fig: exo-disc2}). At large masses 
(20--30~\mearth), the frequency of class~II discs (black curve) is
sufficient to explain the incidence rates for debris discs and gas
giants in the constant core mass (cyan curve) and the variable core
mass (green curve) models. Within this group, somewhat larger core 
masses for gas giants (up to 20~\mearth) and larger maximum masses
within the parent bodies of debris discs (up to 200~\mearth) still
fit within the available mass reservoirs in the most massive class~II 
discs. At smaller masses, however, class~II discs cannot explain the
high frequency of planets with masses $\le$ 5--10~\mearth. 

The high frequency of microlensing planets is responsible for the large 
discrepancy between the mass reservoir required for the known exoplanets 
and the mass available in class~II discs. 
In a variable core mass model with no 5--30~\mearth\ microlensing
planets ($f_{se} = f_{nep}$ = 0), class~II discs have enough mass 
to explain the frequency of other known exoplanets
(Figure~\ref{fig: exo-disc2}, magenta curve). 
Despite the failure of class~II discs to provide enough mass for the 
known exoplanets, the class I discs 
have more than enough mass 
(Figure~\ref{fig: exo-disc2}, violet curve), with 
roughly 3 times the mass required for the
known exoplanets.

\section{Discussion} 
\label{sec: discuss}

Our analysis demonstrates that the solids contained in the single
star class~II discs of Taurus-Auriga are sufficient to explain the
incidence rates for some, but not all, exoplanet populations. 
The Taurus-Auriga
discs have enough mass for Kepler planets ($P\le400$\,d), 
giant planets ($a \le20$\,AU), and debris discs ($>20$\,AU).
However, published single star class~II disc masses are insufficient 
to account for the large observed incidence rates of 
5--30~\mearth\ microlensing planets (0.5--10\,AU). 
Thus, the mass of solids locked up in planets and 
the parent bodies of debris discs 
exceeds the reservoir of solids in the protoplanetary discs commonly 
assumed to be the starting points of the planet formation process. 

In the next sections, we discuss ways to resolve the discrepancy
between the observed masses in exoplanets and single star class~II discs.
After considering possible caveats in our analysis (\ref{sec: caveats}), 
we address factors that probably widen the gap between the two mass 
reservoirs (\ref{sec: factors}). We then describe our preferred 
solution for reconciling the mass budget of protoplanetary discs
(\S\ref{sec: epoch}).

\subsection{Caveats}
\label{sec: caveats}

There are three main uncertainties in our analysis. 
Firstly,
the microlensing incidence rates have much larger uncertainties 
than the modest errors in the rates for debris discs and 
the HARPS and \kep\ planets. 
The microlensing results indicate 
a clear preference for an incidence rate larger than 0 
for 5--30~\mearth\ planets  
\citep{gould2006,cassan2012}. 
The good agreement between the giant planet ($>$ 100~\mearth) 
incidence rates from microlensing and radial velocity 
measurements is also encouraging.  
However,
the 3$\sigma$ error bars for 5--10~\mearth\ and 10--30~\mearth\ 
planets formally allow any incidence rate between 0 and 1. 

In the Monte Carlo simulations of \S\ref{sec: montecarlo}, we
demonstrate that a negligible incidence rate for 
5--30~\mearth\ microlensing planets enables a good match between
the available mass in single-star class~II discs and the mass 
observed in debris discs at $\ge$ 20~AU, gas giants at 
$\le20$~AU, 
and 
\kep\ planets at $\le$ 1~AU. Simulations with incidence rates of
10\% for 5--10~\mearth\ and for 10--30~\mearth\ planets 
($f_{se} + f_{nep} \le$ 0.2)
at 1--20~AU also yield reasonable matches with the class~II disc 
mass distribution. For larger incidence rates 
($f_{se} + f_{nep}$ $\ge$ 0.2), 
the match is poor. 
Revised  
incidence rates with a factor of two reduction in the 1$\sigma$ 
error bars would enable a much more accurate assessment of the
demand placed on the disc solid mass budget by 
5--30\,\mearth\ planets.

Secondly, we have implicitly assumed the core accretion picture 
of planet formation, where ice and gas giants have significant 
solid cores \citep{pollack1996}. In the disc instability picture
\citep[e.g.,][]{boss2000,boss2005}, gas giants form in a 
gravitationally unstable disc and then migrate close to the
host star. The viability of this mechanism is uncertain
\citep[e.g.,][]{raf2005,clarke2009a,rice2009,krat2010,helled2013}.
However, if disc instabilities 
commonly produce ice and 
gas giants, our analysis overestimates the solids contained in 
exoplanets.
 
To pursue this idea in the Monte Carlo simulations, 
we define $f_{ca}$ 
as the fraction of ice and gas giants formed with massive solid
cores as in the core accretion model. Disk instabilities must 
then produce a fraction $1 - f_{ca}$ of ice and gas giants. 
Although ice and gas giants 
formed by disc instability may have modest solid cores 
\citep{helled2013}, we assume for simplicity that 
they
make no contribution to the
mass distributions of solids in exoplanets. 
The fraction of $>$10~\mearth\ microlensing planets 
with massive solid cores is $f_{ca} f_{nep}$ for Neptunes 
and $f_{ca} f_{gg}$ for gas giants. Repeating the Monte Carlo
simulations (\S 4.2) with a variable $f_{ca}$ allows us to measure the 
importance of the formation pathway in setting the exoplanet mass 
distribution. 

Even with a negligible fraction of ice and gas giants from
core accretion, it is impossible to eliminate the discrepancy
between the Monte Carlo mass distribution of exoplanets and 
the single star class II disc mass distribution. For any $f_{ca}$, 
the large incidence rate for 5--10~\mearth\ microlensing planets
precludes a match at small masses, 
below 5~\mearth.  With
$f_{ca} \le 0.1$, the Monte Carlo results match the observed
disc masses 
above 5~\mearth. Much larger fractions of 
giant planets from core accretion (e.g., $f_{ca} \ge$ 0.2) 
yield poor matches 
in the interval 5--10~\mearth. 

Although we focus on several specific approaches to reduce the
mass distribution of exoplanets to the level of the Taurus-Auriga
class~II sources, some combination of (i) a smaller incidence 
rate for microlensing planets, (ii) a smaller core mass for planets
formed by core accretion, and (iii) a smaller fraction of planets
produced by core accretion yields similar results. 
Among these options, reducing the incidence rate for 
5--10~\mearth\ microlensing planets is essential. 
Lowering the incidence rate for 10--30~\mearth\ microlensing
planets can be combined with the details of the formation
mechanism to yield a better match at 5--10~\mearth.

Finally, the disc masses we use may be underestimated. In contrast
to the simple scaling between submillimeter flux and disc mass
employed here (eq. [4]; Andrews et al.\ 2013 and S.\ Andrews 2014,
private communication), a relatively small subset of discs (the
brighter submillimeter population) also have mass estimates based
on more sophisticated fits to broad-band SEDs and continuum
visibilities \citep[e.g.,][]{and2009,and2010,and2011}. In 
approximately half of the
systems, the more sophisticated disc masses are larger than the optically
thin estimates by factors of 3-8. If these results were to apply
broadly to all Taurus-Auriga discs, our disc masses for single class
II sources must be increased by a factor of $\sim 2-3$. 

However, more sophisticated analyses of the complete ensemble of
Taurus-Auriga discs may lead to little change in the median disc
mass. 
Among sources studied to date, disc size generally increases
with submillimeter luminosity \citep{and2010}. 
Larger, brighter discs are therefore cooler than average, requiring
more dust to produce the same submillimeter flux. They are therefore
expected to have the {\it largest} corrections to the disc masses
derived from the simple optically thin relation (eq. [4]). If the
trend of disc size with submillimeter luminosity also applies at
very low submillimeter fluxes, 
the 
fainter discs are smaller than
average and they will tend to have smaller masses than eq.~(4) predicts.
The true disc mass distribution may then be broader than the
distribution derived from eq.~(4), extending to both higher 
and lower disc masses, without altering the median disc mass. Fits 
to SEDs and visibilities from interferometric observations of the 
faint disc population are needed to address this issue.

\subsection{Exacerbating Factors}
\label{sec: factors}

While the factors described above can reduce the high frequency of 
disc solid masses in the 1--20~\mearth\ range, 
other issues probably increase it.

The ensemble of known planets with $\Mp \le$ 5~\mearth\ is 
incomplete.  For radial velocity and transit observations, selection 
effects and sensitivity 
limit the population of planets with $\Mp \le$ 1--2~\mearth\ at 
$a \le$~1~AU \citep[e.g.,][] {mayor2011,youdin2011b,dong2013,burke2014}. 
The sensitivity of published microlensing observations precludes planet 
detections for $\Mp \le$ 5~\mearth\ at 1--10~AU \citep[e.g.,][]
{gould2006,gould2010,sumi2011,cassan2012}. Because the number of planets 
grows rapidly with decreasing mass 
down to 5~\mearth\
(Figure 
\ref{fig: exo-mass1}), we expect that enhancing the sensitivity for
microlensing, radial velocity, and transit observations would enable 
many detections of lower mass planets. Increasing the population of
low mass planets in our analysis would add to the discrepancy between
the mass in exoplanets and the disc masses in single class~II sources.

The population of exoplanets beyond 20~AU is also incomplete. This
region is inaccessible to microlensing, radial velocity, and transit 
techniques.  Robust detections from direct imaging are limited to 
massive planets orbiting young host stars 
\citep[e.g.,][]{marois2008,lagrange2010,kraus2012,currie2014}.
Analyses of existing detections and null results from extensive
imaging surveys of solar-type stars suggest maximum incidence rates 
for gas giants of $\sim$ 10\% at $a \ge$ 20~AU 
\citep{nielsen2010,janson2012a,biller2013,yama2013}. 
If gas giants are roughly 
half as common as debris discs beyond 20~AU, it is much harder for
single class II sources to have enough mass for planets at smaller
semimajor axes.
Direct imaging technology on the largest ground-based telescopes is
improving rapidly \citep[e.g.,][]{hugot2012,jova2013,mac2014}.  New 
exoplanet imagers (GPI, SCExAO, and SPHERE) should enable better
estimates for incidence rates of gas giants around young stars. 
As these estimates improve, it will be easier to assess the impact
of exoplanets beyond 20~AU on the mass budget of the exoplanet population.

As discussed in \S\ref{sec: discs}, the single star class~II disc 
mass distribution excludes the class~III sources.  Comprising 
roughly 25\% of the T Tauri stars in Taurus-Auriga, the
class~III sources have little or no evidence for large reservoirs
of solids \citep{and2005}.  Including these discless stars in the 
class~II disc mass distribution greatly reduces the frequency of 
T Tauri discs with enough solids to produce the known exoplanets 
(Figure~\ref{fig: disc-mass}, red curve). 

This argument assumes that class III objects are similar in age to
class II objects, despite the difference in their evolutionary state.
An overlapping age distribution for class II and III objects seems 
reasonable for Taurus-Auriga, where the class IIs are intermixed with 
the class IIIs in the HR diagram \citep{cohen1979,kh1995,gudel2007}.
The spatial distributions of class II and class III sources are also
similar \citep{luhman2010}. Although commonly accepted today, 
future observations may alter this picture. An HR diagram for all 
of the sources in the new, more complete Luhman et al.\ (2010) sample 
may show that a signficant fraction of the class III objects are older 
than the average class II object. Similarly, parallax distances measured 
with Gaia may reveal that class III objects are closer than the class 
II objects and therefore older, 
on average \citep[e.g.,][]{bertout2007}. In assuming that class II 
and III objects have similar ages, our discussion here represents 
a worst case scenario.

However, 
the disposition of the class~III sources depends on our
uncertain knowledge of their initial state.  If class~III
sources formed with little or no mass in a disc, then their 
weak submillimeter emission is intrinsic. It is then
appropriate to add their negligible solid masses to the
class~II mass distribution, widening the discrepancy 
between the `initial' mass distribution of single 
class~II+III sources and the `final' mass distribution 
of exoplanets.

Alternatively, class~III sources could have formed with
substantial disc masses and evolved rapidly into their
current discless state. If this evolution converted 
small grains into planetesimals or planets,
then it is appropriate to augment the class~II mass
distribution with the `original' disc masses of the
class~III sources. If these original disc masses 
exceed the median disc mass, then class~III sources
reduce the discrepancy between the initial and final
mass distributions.

Searches for reservoirs of solids in the class~III sources
would illuminate their impact on the initial mass reservoirs
of T Tauri discs.
Although identifying planetesimals in class~III sources
seems unlikely, measuring the fraction with debris discs 
or planets would place clear limits on the reservoir of 
solids orbiting these young stars. 
As discussed in \citet{cieza2013}, current limits on 
the IR excesses of WTTS and class III objects place only 
limited constraints on the mass in solids that 
may reside in these systems. 

Another issue is the efficiency with which disc solids are 
converted into planetary systems. 
The solid mass reservoirs in the known exoplanets clearly
represent a {\it lower limit} on the solids initially available in
protoplanetary discs.  Observations of warm dust at radial
distances of 0.5--2~AU from many solar-type stars suggest 
a clear inefficiency in terrestrial planet formation
\citep[e.g.,][]{currie2007b,rhee2008,melis2010,chen2014,sier2014}. 
Numerical simulations
suggest plausible inefficiencies of 10\% to more than 25\%
\citep[e.g.,][]{weth1993,agnor2004,kb2004b,asphaug2006,raymond2011}.
If such inefficiencies are typical of the formation of ice and
gas giants, initial disc masses must be correspondingly larger.

The large eccentricities ($e$) of gas giants, Neptunes, and 
super-Earths from the \kep\ and radial velocity data imply
even larger inefficiencies.
Approximately half of the giant planet population 
($M>0.2 M_J$) have orbits with $e >0.2$ \citep{howard2013}. 
Super-Earths and Neptunes ($M < 30-40~\Mearth$) have 
eccentricities up to $\sim 0.45$ (Mayor et al.\ 2011). 
To produce such large $e$, current theoretical studies favor 
scattering among massive protoplanets or planets, where 
interactions between two or more massive planets 
eject\footnote{In some calculations, the 'ejected' star is
placed on an eccentric orbit with large $a$. Because this planet
is not included in any of our planet reservoirs, it is lost
to our census.} one
planet from the system and place another on an eccentric orbit 
close to the host star \citep{rasio1996,juric2008,raymond2010}. 
If this picture is correct, the initial masses of these planetary
systems must have been 1.5--2 times more massive than observed 
today.
If some planets are lost because they migrate into their host 
stars \citep[e.g.,][]{trill1998},
the initial masses must be even larger.

Dynamical models of the formation of the Solar System  
suggest a similarly
large inefficiency. Assuming core masses of 5--10~\mearth\ for 
each gas giant and 2--5~\mearth\ for the terrestrial planets,
the total mass in solids contained in the planets is roughly 
25--45~\mearth.  Recent numerical simulations require an 
additional 30~\mearth\ of solids to stabilize the orbits of 
the gas giants in their current architecture 
\citep[e.g.,][and references therein]{morbi2013}.
This Nice model therefore requires a minimum initial solid 
disc mass of roughly 55--75~\mearth\ 
\citep[see also][]{desch2007,dawson2012} -- about twice 
the mass of the planets -- to produce what we think of as 
a `typical' planetary system without a substantial debris disc. 

\begin{figure}
\centering
\includegraphics[width=15cm]{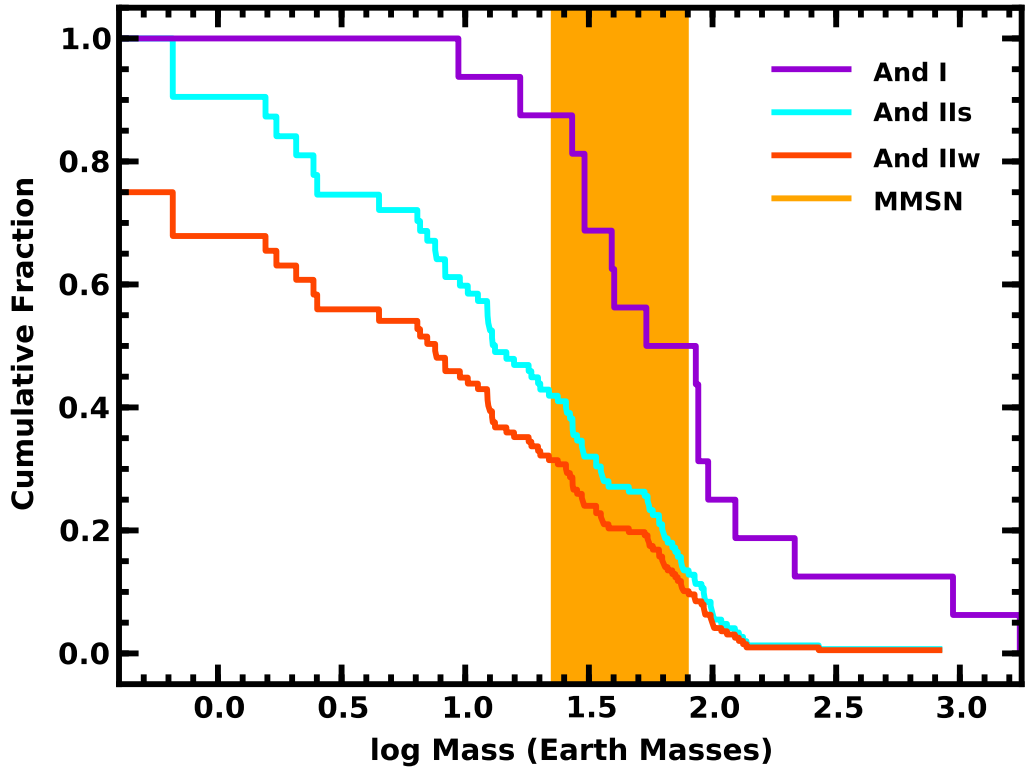}
\caption{
Comparison of the minimum mass solar nebula (MMSN) and 
`minimum mass exoplanet nebulae' (MMEN) with the mass
distributions of protoplanetary discs.
The orange bar indicates the mass range for the MMSN and MMEN.
Solid lines plot the mass distributions of the discs in 
class I protostars (`And I`, upper violet curve),
class~II sources (`And IIs`, middle cyan curve), and
class~II+III sources (`And IIw`, lower red curve).
The MMSN and MMEN lie well above the median disc mass 
of the class II and class II+III distributions.
\label{fig: exo-disc3}
}
\end{figure}

Figure~\ref{fig: exo-disc3} 
illustrates the impact
of inefficient planet formation.  More than 40\% of the class 
II discs (30\% of class~II+III discs) have masses larger than
the most minimal MMSN 
(25 \mearth). 
This minimal MMSN is therefore fairly typical. 
However, allowing for the maximum core masses of the gas 
giants and the inefficiency of the Nice model, fewer than 
15\% of the class~II discs have sufficient mass 
(75 \mearth) 
to produce 
the planets in the Solar System. The Solar System is then
very atypical.

A similar picture arises in models for the formation of the known 
exoplanets \citep[e.g.,][]{ida2008a,ida2008b}.
In their studies of 
exoplanets within 1\,AU of their host stars,
\citet{hansen2012,hansen2013} and \citet{chiang2013}
explore the disc surface density distributions required to 
build the known exoplanet populations {\it in situ}. 
\citet{raymond2014} consider whether the exoplanet populations 
form at larger distances and then migrate inward to their current locations.
In either approach, the masses required for these `minimum mass
exosolar nebulae' are within the range outlined for the MMSN in
Figure 6. 
The large masses invoked in these scenarios draw from
the upper 10-15\% \citep{hansen2012,hansen2013,raymond2014}
or upper 35\% \citep{chiang2013} of class II disc masses.

\subsection{When Does Planet Formation Begin?}
\label{sec: epoch}

One way to resolve the discrepancy between the solid masses in 
planetary systems and single star class~II discs is to consider 
the  possibility that the submillimeter disc masses are not 
``primordial'' \citep[e.g.,][]{greaves2010b,vorobyov2011,will2012}.
If the dust distribution 
evolves significantly on time scales of a few Myr, 
current measurements underestimate the true solid 
reservoirs available for planet formation. Because 
the inner disc is optically thick, concentrating 
mm--cm-sized solids in the inner disc hides them 
from submillimeter telescopes.  Alternatively, 
growing solids to km or larger sizes throughout 
the disc prevents detection. Either explanation 
implies that more than 50\% of the solids in 
class~II discs is already locked up in large 
particles or is sequestered at small disc radii
(or both).

Current data allow either possibility. If small solids
are sequestered inside a few AU, then the disc has a
steep surface density distribution, $\Sigma \propto a^{-p},$
with large $p$. 
Gas drag and
other physical processes naturally concentrate grains
in the disc \citep{naka1986,youdin2004a,brauer2008,
birn2010,birn2012,laibe2012,pinte2014},
leading to configurations with 
$p \ge 1.5$ \citep{birnstiel2014,laibe2014}.
Observationally, the value of $p$ is not well constrained. 
Although initial interferometric
observations were interpreted as evidence for 
$p \approx$ 1 for some T Tauri
discs \citep{and2010}, 
subsequent multiwavelength studies 
\citep{perez2012} 
demonstrate real variations in the grain size distribution 
with radius that 
make it difficult to constrain $p$ with existing observations 
(S. Andrews, private communication).

The growth of solids throughout class~II discs is also plausible. 
Inside 30~AU, the time scale for the growth and radial drift 
of cm-sized particles is short \citep{raf2003d,youdin2004a,rice2004,
rice2006,clarke2009b,birn2010,birn2012,windmark2012,garaud2013,laibe2014}.
Some $\sim 90$\% of the small solid particles in the 
inner 30\,AU can be converted into cm-sized or larger 
pebbles on timescales of $1 - 3 \times 10^4$\,yr
\citep{birn2010, birn2012}.
Although agglomeration into larger and larger objects may be possible 
\citep[see][and references therein]{rice2006,clarke2009b,windmark2012,garaud2013},
recent studies focus on models where small solid particles
drift radially inward and concentrate in local pressure 
maxima \citep[e.g.,][and references therein]{dittrich2013}.
When the local gas-to-dust 
ratio reaches values above unity, streaming instabilities 
concentrate pebbles into aggregates which collapse 
gravitationally into much larger planetesimals \citep[e.g.,][]
{youdin2005,johansen2007,johan2009,youdin2011a}. These
planetesimals rapidly accrete the pebbles, evolving into
1--10~\mearth\ protoplanets on time scales $\le$ 1--3~Myr,
comparable to the ages of Taurus-Auriga T Tauri stars
\citep[e.g.,][]{raf2005,ormel2010a,bk2011a,lamb2012,chambers2014}.

There is independent evidence for rapid planetesimal formation 
in the Solar System, based on radiometric analyses of meteorites 
\citep[e.g.,][]{bizzarro2005,
kleine2009,schulz2009,dauphas2011a,dauphas2011b,sugiura2014}.
The elemental abundances of the oldest solar system objects -- the
mm- to cm-sized calcium aluminum inclusions (CAIs) -- indicate 
that these objects formed on time scales $\le$ 0.1~Myr during a period 
when the Sun emitted copious high energy particles and x-rays. 
Studies place the epoch of CAI formation during the late
class I or early class II phase \citep{dauphas2011a}, 
with CAIs accumlating into differentiated 
planetesimals in the next $\sim$ 1--10 Myr 
\citep{kleine2009,dauphas2011a}.
Thus, the meteoritic record implies that planetesimal
formation was well underway at the onset of the class II phase 
of solar system history.

Our conclusions expand on previous results derived from
comparisons of the disc and exoplanet mass distributions
\citep{greaves2010b,greaves2011,vorobyov2011, will2012}.  
In earlier studies, both of the adopted mass distributions were incomplete
at low masses; thus, the studies focused on the ability of the most
massive discs to produce the gas giant planets \citep[e.g.,][]{cum2008}.
Because gas giant planet formation is predicted theoretically to
be inefficient in concentrating disc material into planets
\citep[e.g.,][]{dod-rob2008,dod-rob2009}, giant planets are expected
to form only in discs with masses $\ge$ 2--3 times the mass of the MMSN.
With few such discs among class II sources, \citet{greaves2010b}
inferred that planet formation begins at an earlier epoch when discs
are more massive \citep[see also][]{vorobyov2011,greaves2011,will2012}.

This inference relies on the assumed (in)efficiency of planet
formation, a quantity that is not well known 
\citep[e.g.,][]{armitage2003,ida2008a,ida2008b,dod-rob2009,fogg2009,
coleman2014}.  
However, as we have
shown, the improved statistics for disc and exoplanet masses now allow
us to bypass this difficulty.  The new data directly imply a mismatch
between the masses in 1--10~\mearth\ planets and the masses available
in class II discs (\S4). Thus, these data lead to the stronger
implication that planet formation is underway in class II discs
{\it independent of assumptions about the efficiency of planet
formation.  }

This discussion suggests that primordial disc masses are probably 
much larger than the disc masses of single class~II sources 
measured by \citet{and2013}. With masses 2--5 times larger,
the discs in class~I sources contain more than enough mass to
explain the masses of exoplanets (Figure~\ref{fig: exo-disc3})
and 
to accommodate a factor of $>2$ inefficiency 
in converting disc solids into planets 
\citep[\S5.2; see also ][]{greaves2011}.
The cumulative disc mass distribution of the ``evolutionarily younger''
Class I sources is steeper than that of the Class~II sources and
roughly parallels the Monte Carlo mass distribution derived 
from the incidence rates for exoplanets (Figure~5). In this picture,
Taurus-Auriga protoplanetary discs can be the precursors 
of exoplanet systems if Class I objects evolve into Class~II 
objects while preserving a large fraction of their disc solids.

Achieving this outcome may be challenging.  
Throughout the class~I 
and class~II phases, stars continue to accrete mass from their 
discs and to eject disc mass in winds and jets 
\citep[e.g.,][]{naj1994,hart1996,hart1998,offner2011,
frank2014,turner2014}.
If accretion and mass loss remove a significant fraction of the 
solids from a class~I disc, then the disc will not contain enough
material to produce the known exoplanets. However, the time scales
for agglomeration and radial transport of solids through the disc 
are much shorter than the time scales for accretion and mass loss.
Thus, it seems plausible that known physical processes can concentrate 
the solids into large planetesimals while gas accretes onto the
central star and is ejected in jets and winds.

Once planetesimals form, they may follow several evolutionary
paths. Two scenarios plausibly explain the \kep\ planets:
(i) inward migration followed by 
{\it in situ} growth into protoplanets inside 1~AU 
\citep{hansen2012,chiang2013} or 
(ii) growth into protoplanets at several AU followed by inward 
migration \citep{ida2008a,chambers2008,raymond2011}.
Growth at several AU with little or no migration is necessary 
to explain the microlensing planets \citep{ida2005,ida2008a,
raymond2011,raymond2014}. Finally, negligible radial transport
followed by protoplanet growth beyond 10~AU is responsible for
most debris discs. Our results suggest that the first step in
any of these scenarios -- the formation of planetesimals -- is
already underway in the Taurus-Auriga class~II discs.

\section{Summary and Conclusions} 

The emerging paradigm of planet formation as a common outcome 
of T Tauri disc evolution 
sheds new light on when planet formation begins and/or 
the efficiency with which it occurs. 
We have compared the solids present in known exoplanetary systems 
(exoplanets and debris discs) with the solid reservoirs 
reported for T Tauri discs, the presumed birthplaces of 
planets (\S2 and \S3). 
For the comparison, we used a simple tally approach that considers
the solid masses available in the outer ($>20$\,AU), inner ($< 4$\,AU),
and middle (4--20\,AU) regions of the disc, compared with the demands
on these solid reservoirs placed by debris discs, Kepler planets,
and the remaining planet populations, respectively (\S 4.1). We
also used a Monte Carlo approach to create ensembles of systems
with planets and debris discs, based on their known incidence rates,
and compared the solid mass distribution of the ensembles with that
of protoplanetary discs (\S 4.2). The latter approach only considers
the total mass budget of each system and ignores the fraction of
the disc mass at different disc radii that might be available to
form the planet and debris disc populations.

In both approaches,  
the solids in single-star class~II discs are 
barely
adequate to account for the solids contained in most of the known 
populations of planets (giant planets within 10\,AU, 
super-Earths and Neptunes within 400\,d) and debris discs. 
Moreover, the 5--30\,\mearth\ population 
of planets 
at 0.5--10\,AU discovered 
by microlensing is too numerous (at their reported incidence rate) 
to explain with the known reservoirs of T Tauri disc solids (\S4). 
The discrepancy between the solid mass budgets of planets and 
class II discs implies that planet formation is already underway in the 
class II phase. 

There are three main uncertainties in our analysis: 
(i) large errors in the incidence rates for 
5--30~\mearth\ microlensing planets, 
(ii) uncertain solid core masses for 
Neptunes and gas giants at 1--20~AU, and 
(iii) the reliability of disc masses from the simple optically thin
estimate (eq.\ [4]). Lower incidence rates, smaller core masses, and
larger disc masses would reduce the shortfall.
While a clear preference for a significant microlensing rate is
already apparent, the incidence rate would have to be reduced by a
factor of 5--6 (from 52--62\% to 10\%) to eliminate the discrepancy. 
Further work to refine the incidence rate of the microlensing 
population would therefore be valuable to confirm   
the apparent shortfall. 

The observed correlation between planet incidence rate 
and stellar metallicity appears to favor core accretion as 
the dominant pathway for the radial velocity planet population 
within a few AU \citep{fischer2005,mayor2011,miller2011}. 
Thus, the large core masses we adopt seem reasonable.
However, 
a similar analysis of the properties of the (5--30~\mearth, 
1--10\,AU) microlensing planet population 
would be valuable in determining
whether 
the smaller solid core mass expected from the disc 
instability mechanism 
is preferred for planets at larger orbital radii. 

Despite these uncertainties,
our comparison of the solids in the known planets and 
single-star class II discs is fairly conservative.  Other 
factors are likely to broaden the gap in the solid mass budget 
between discs and planets. Our comparison ignores the population 
of discless class III sources, the likely inefficiency of planet 
formation (factor $> 2$), and planet populations yet to be discovered 
(\S5.2).

For these reasons,  
it seems difficult to escape the conclusion that 
planet formation is already underway in class~II discs. 
This conclusion can be avoided only if the microlensing planets make a
negligible contribution to the solid mass budget and none of the
exacerbating factors (\S 5.2) play a significant role. 
If this extreme set of conditions is true, we would still conclude 
planet formation must be extremely frugal---all existing solids 
in class~II discs will be turned into planets without loss. 
If this is the case, it is important to reconsider planet formation scenarios
that are profligate in their use of disc material and rely on {\it
only} the most massive class II discs to form known exoplanets.
Because it does not require extreme assumptions, it seems more
likely that the original interpretation is correct: planet formation
begins early, and the comparison between the solids in class~II
discs and in known planetary systems provides a clear constraint
on the epoch of planetesimal and planet formation.

\section*{Acknowledgments}

We thank Sean Andrews, Til Birnstiel, Ben Bromley, Margaret Geller,
John Johnson, and Hubert Klahr for valuable discussions and comments
on the manuscript.
We also thank the referee for a helpful report that improved the 
clarity of the manuscript. 
Portions of this project were supported by the 
{\it NASA Astrophysics Theory}
and {\it Origins of Solar Systems} programs through grant NNX10AF35G, the 
{\it NASA Outer Planets Program} through grant NNX11AM37G, and by 
the Institute for Theory and Computation at the Harvard-Smithsonian Center 
for Astrophysics. 


\bibliography{ms.bbl}

\clearpage

\clearpage

\clearpage

\clearpage

\clearpage

\clearpage

\clearpage

\bsp

\label{lastpage}

\end{document}